\documentclass[twocolum n, superscriptaddress, pre, aps]{revtex4-1}
\usepackage{amsmath}
\usepackage{amssymb}
\usepackage{amsfonts}
\usepackage{bm}
\usepackage{amssymb}
\usepackage{graphicx}
\usepackage{epsfig}
\usepackage{dcolumn}
\usepackage{txfonts}
\usepackage{makeidx}
\usepackage{color}
\usepackage{mathtools}
\usepackage{threeparttable}
\usepackage[colorlinks,linkcolor=red,anchorcolor=blue,citecolor=green]{hyperref}

\begin{document}
%\title{Entanglement Entropy in Tensor-Train RNN for High-order Nonlinear Dynamics}
\title{Tensor-networks for High-order Polynomial Approximation: An Many-body Physics Perspective}

\author{Tong Yang} 
\affiliation{Department of Physics, Boston College, Chestnut Hill, Massachusetts, USA}
\affiliation{Department of Computer Science, Brandeis University, Waltham, Massachusetts, USA}
\affiliation{Center for Polymer Studies, Boston University, Boston, Massachusetts, USA}
\affiliation{Machine Learning Center of Excellence, J.P. Morgan, Kowloon, Hong Kong}

\begin{abstract}
	We analyze the problem of high-order polynomial approximation from a many-body physics perspective, and demonstrate the descriptive power of entanglement entropy in capturing model capacity and task complexity.
	Instantiated with a high-order nonlinear dynamics modeling problem, tensor-network models are investigated and exhibit promising modeling advantages.
	This novel perspective establish a connection between quantum information and functional approximation, which worth further exploration in future research.
\end{abstract}

\maketitle

\section{Background Introduction}

Machine Learning (ML) has draw tremendous attention in modeling complex relations among high dimensional variables, a significant advantage of which is the model capacity of deep neural networks (DNN).
From an approximation perspective, DNNs capture interactions among different variable dimensions by stacking layers, which effectively form higher order tensors.
To approximate tensors efficiently, however, there is another set of models: tensor networks (TN).

TNs are proposed and implemented in various disciplines, including tensor approximation, quantum information, condensed matter theory, etc., and are especially useful to capture many-body physics with a large set of degrees of freedom.
This scenario is similar to many ML tasks, both resolving the challenge of interactions among different variables.

Inspired by this similarity, in this work we propose to implement TNs on the problem of high-order polynomial approximation (HPA).
A typical case where HPA works would be high-order nonlinear dynamics (HND), where the system evolution is governed by certain high-order polynomial functions of a set of state variables.
The polynomial order depends on both the variable order in evolution functions and the order of derivatives in differential equations.
We would deliver a discussion on TNs for HPA, taking HND modeling as the instantiating scenario, partially because it is more straightforward to connect with some existing works.
However, it is quite straightforward to generalize the discussion to arbitrary HPA problems.
Interestingly, we would analyze the problem from a many-body physics perspective, which then establish a connection between entanglement and function approximation.
This novel perspective would shed new light on general modeling tasks.

\section{Problem Formulation}\label{1}
We firstly would define the concerned HND problem in an exact way. We would consider a RNN form of the problem, and introduce the Hilbert space of the problem in RNN formulation.

\subsection{General Problem Frame}\label{1a}
Considering a system whose state at time $t$ can be fully captured by a high-dimensional vector:
\begin{align}
\mathbf{x}_t \in \mathbb{R}^{\chi},
\end{align}
where $\chi$ is the dimension of state-vectors. The most general dynamics governing the system can be formulated as:
\begin{align}
\mathcal{F}\bigg[\mathbf{x}_t^{(0)}, \mathbf{x}_t^{(1)}, \cdots \mathbf{x}_t^{(d)}; \xi\bigg] = 0,
\end{align}
where $\mathbf{x}_t^{(i)}$ represents the $i$-th order derivative of state-vector, and $\mathcal{F}$ is a nonlinear function, which usually represents a continuous dynamical system. There are two types of nonlinear dynamics: continuous and discrete ones. In this work, to simplify the question for a general modeling perspective, we would concentrate on discrete dynamics. In discrete case, where differential equations are replaced by difference equations, it is legitimate to rewrite the above dynamics as:
\begin{align}
\mathbf{x}_{t+1} = \mathcal{F}\big[\mathbf{x}_t, \cdots \mathbf{x}_{t-\tau}; \xi\big].
\end{align}
The result can be also generalized to continuous systems as long as no singularity encountered approaching continuous limit, which in general is hard to justify though.

From a modeling perspective on the forecasting problem, the general goal is to find a proper mapping $F$, such that:
\begin{align}
F: (\mathbf{x}_1, \mathbf{x}_2, \cdots, \mathbf{x}_t) \mapsto (\mathbf{x}_{t+1}, \mathbf{x}_{t+2}, \cdots, \mathbf{x}_{t+T}).
\end{align}
More specifically, we could formulate this into a parameter-searching (optimization) problem: $F \equiv F_{W}$, where $W$ represents parameters that we are targeting at. For simplicity, we would rewrite forecasting the problem as:
\begin{align}\label{general_prob}
\mathbf{x}_{t+1} = F_{W}\big[\mathbf{x}_1, \mathbf{x}_2, \cdots, \mathbf{x}_t\big].
\end{align}
In the context of "learning", we would be given some history information to feed into the searching process, and find the optimal parameters $W$.

\subsection{RNN Form of HND Forecasting}\label{1b}
As mentioned earlier, we would apply a RNN model to the problem. It has been proposed in Yu (2017)\cite{yu} that a RNN model for a general HND problem can be formulated as an iterating structure on hidden units $\mathbf{h}_t$ :
\begin{align}
[\mathbf{h}_t]_{\alpha} = F\bigg[\mathcal{W}_{\alpha}\mathbf{x}_t + tTr\bigg(W_{\alpha}^{i_1\cdots i_P}\mathbf{s}^{_{(t-1)}}_{i_1}\otimes\cdots\otimes\mathbf{s}^{_{(t-1)}}_{i_P}\bigg)\bigg],
\end{align}
where a "history vector" is defined as: 
\begin{align}
\mathbf{s}^{_{(t-1)}} = \big[1, \mathbf{h}_{t-1}, \cdots, \mathbf{h}_{t-L}\big],
\end{align}
representing a $L$-lag process; and $tTr(\cdot)$ represents a tensor-trace contracting all indexes $\{i_1, \cdots, i_P\}$. Note that we have $P$ identical copies of $\mathbf{s}^{_{(t-1)}}$ variable, which codes higher-order elements as:
\begin{align}
\prod_{m=1}^{L}\mathbf{h}_{t-m}^{p_m}, \qquad s.t. \bigg(\sum_{m=1}^Lp_m \leq P\bigg),
\end{align}
and thus can compose all possible polynomials upto $P$-th order. The state-vector $\mathbf{x}_{t+1}$ can then directly be obtained from $\mathbf{h}_t$ thorough another function $\sigma(\mathbf{h})$ as in general RNN (e.g. Yu\cite{yu}).
Due to the specific concern of this work, we would care more about result from historical information, hence, more on parameters $W_{\alpha}^{i_1\cdots i_P}$, while ignore both $\mathcal{W}$ and $\sigma$ for now. This is still a reasonable analysis since the optimization procedure in principle can be viewed separately on different parameters and then reach the final result through one more "combined optimizing" stage, and the current concern can then be viewed as an analysis restricted to the first stage. In fact, the representing power of RNN for long-history is indeed majorly determined by $W$ itself.

Without loss of generality, we could consider the case where $x$ and $h$ are scalars, and the higher-order analysis follows naturally. In this case, we would focus on the following "historical impact" part within the high-order RNN structure:
\begin{align}\label{PL-prob}
\psi_W &= tTr\bigg(W^{i_1\cdots i_P}s_{i_1}\otimes\cdots\otimes s_{i_P}\bigg) \nonumber \\
s_i &\in \big[1, h_{t-1}, \cdots, h_{t-L}\big].
\end{align}
The above function $\psi_W$, whose forms is determined by $W$, is the problem defined for our current work, which contains both long-history correlations and higher order dynamics. In the following we would call this an \textit{order-$P$ lag-$L$ problem}.

\subsection{Hilbert Space}\label{1c}
For the purpose of a rigorous analysis, we would define the Hilbert space for an order-$P$ lag-$L$ problem. Firstly, we map the formulation \eqref{PL-prob} into a spin model with spin amplitude $S=\frac{L}{2}$ ($L$'s parity can be safely ignored). More precisely:
\begin{align}
\mathbf{s}_p &\leftrightarrow \mathbf{S}_p \nonumber \\
s_{i_p} &\leftrightarrow S_p^z\nonumber \\
\big[1, h_{t-1}, \cdots, h_{t-L}\big] &\leftrightarrow \big[-S, -S+1, \cdots, S\big],
\end{align}
and then the "historical impact" part \eqref{PL-prob} can be viewed as a many-body wave-function in condensed matter physics:
\begin{align}\label{spin-wf}
|\Psi_W\rangle = \sum_{\{i_1,\cdots i_P\}} W^{i_1\cdots i_P}|i_1\rangle\otimes\cdots\otimes|i_P\rangle.
\end{align}

From this mapping, it is clear that the Hilbert space is expanded by the following basis:
\begin{align}\label{Hilbert0}
\mathcal{H}: 
\bigg\{|i_1\rangle\otimes\cdots\otimes|i_P\rangle\bigg\}, \quad \forall i\in \big[1, h_{t-1}, \cdots, h_{t-L}\big],
\end{align}
and the searching problem can be viewed as searching a wavefunction $\Psi_W$ determined by parameters $W$ in the Hilbert space defined above.

Le us verify the legitimacy of defining this Hilbert space. All definitions of Hilbert space can be rigorously satisfied as long as we define following inner product:
\begin{align}
\langle\Psi_{W_1}, \Psi_{W_2}\rangle = \sum_{\{i_1, \cdots, i_P\}} \bigg(W^{i_1\cdots i_P}_1\cdot W^{i_1\cdots i_P}_2\bigg),
\end{align}
which satisfies:
\begin{align}
\langle x, x\rangle &\geq 0; \nonumber \\
\langle x, y\rangle &= \langle y, x\rangle; \nonumber \\
\langle ax_1+ bx_2, y\rangle &= a\langle x_1, y\rangle + b\langle x_2, y\rangle \nonumber.
\end{align}
Thus our definition of Hilbert space is indeed proper. With this in hand, now our problem \eqref{PL-prob} corresponds to a function (or parameter) searching problem in a Hilbert space.

\subsection{Curse of Dimensionality and TT-RNN}\label{1d}

As in most optimization (searching) procedure, the biggest challenge of the above problem is related to the curse of dimensionality. The dimension of the Hilbert space defined above for an order-$P$ lag-$L$ problem is:
\begin{align}
M = (L+1)^P.
\end{align}
Clearly, this result into exponential increase of difficulty for the parameter-searching process. It is reasonable there are systems who has a shorter time ($L$) correlation but higher order ($P$) nonlinear dynamics, in this cases, the exponential increase in $P$ would result into a disaster for computation.

To deal with this, a preceding proposal in Yu (2017)\cite{yu} introduced the so-called Tensor-Train RNN model (TT-RNN), which decomposes the high-rank tensor $W$ into lower-rank ones $A$ in the following way:
\begin{align}\label{TT-RNN}
W^{i_1i_2\cdots i_P} = \sum_{\{\alpha\}}A^{i_1}_{\alpha_0\alpha_1}A^{i_2}_{\alpha_1\alpha_2}\cdots A^{i_P}_{\alpha_{P-1}\alpha_P}.
\end{align}
With tensor-train, they reduced the number of parameters of RNN from $(L+1)^P$ to $(L+1)PR^2$, with virtual dimension $R$ representing the upper bound of virtual index $\{\alpha\}$'s dimension. They claimed the TT-RNN successfully circumvents the curse of dimensionality, which is in sharp contrast to many classical tensor decomposition models, such as the Tucker decomposition.

The above claim, however, is not always promising. Because in general the $R$ might be a function of $L$ and $P$, in which case, the dimension might not be reduced at all. In other words, if one allows a sufficient large value of $R$, say, to infinity, then this decomposition is meaningless. This is actually mentioned in Yu (2017)\cite{yu}, that arbitrary $W$ can be represented by a tensor-train as long as we allow $R\rightarrow\infty$; and a finite value of $R$ would bring in errors. The true question, therefore, is that, given a HND system, if one can use a finite, or even constant independent of $L$ and $P$, value of $R$ to faithfully represent the original RNN model parameter $W$, while keeping the error tolerable.

In fact, from a formal point of view, the reduction of dimension in any case is a nontrivial phenomena, since this would suggest the real problem can be well-approximated in a much smaller low-dimensional subspace. On the one hand, one should note that \textit{it is unreasonable to expect the parameter $W$ of a general HND problem could always be tensor-train-decomposed with finite virtual dimension $R$, and thus be restricted into a smaller subspace}; and in principle, there should always be some problems whose parameter $W$ requires an overall search in the Hilbert space of high dimension (might not be $(L+1)^P$ though). On the other hand, however, the success achieved in previous works, e.g. Yu (2017)\cite{yu} has suggested there are indeed systems whose parameter searching problem is simplifiable.
Therefore, the furthest we may achieve is to:
\begin{itemize}
\item 
classify problems into non-simplifable and simplifiable ones based on certain dimension-related criteria; 
\item
for simplifiable ones, construct proper models, like TT-RNN, to match the subspace where parameters live.
\end{itemize}
This is the major concern of this paper.
In the following, for simplifable problems, we would call a subspace as "effective space" if we can use functions inside to approximate the function (parameters) nicely, and the dimension of this space would be named as "effective dimension". 

Before we close this section, readers should note that, by using other types of decompositions, one may also reduce the number of parameters hence Hilbert space dimension. Therefore the effective space and its dimension is not unique for sure. The problem we are concerning here is based on the proven success of a tensor-train decomposition, which requires a clearer explanation. On the other hand, however, our discussion later may extend beyond the TT-RNN model to more complicated case, while using the same philosophy which is related to the concept of entanglement.

\section{Entanglement Entropy and Tensor Networks}\label{2}
As mentioned above, we are targeting at a problem classification, and related model constructions, with the hint from TT-RNN model.

For the classification, there is a helpful quantity, widely used in both theoretical physics and information theory, that can nicely capture the effective dimension of a given problem: the \textit{entanglement entropy} (EE). In this section, we introduce the concept of entanglement entropy for a general function in the Hilbert space defined for our problem above, whose \textit{scaling behavior} provides us a well-defined criteria to classify functions into different classes. This classification would separate out smaller sub-spaces (effective ones) from the whole Hilbert space. 

With a clear problem classification criteria based on EE scaling, for the model construction, an important class of candidates is Tensor Networks (TN), which have a clear EE scaling behavior that can be relied on for model choosing. We would introduce the generic form of a TN, and also mention its relation to the previously proposed TT-RNN model, which explains the reason behind our criteria choosing.

\subsection{Density Matrix and Entanglement Entropy}\label{2a}

To define entanglement entropy for a wave-function, it is helpful to introduce the concept of density matrix. Firstly, for a pure state described by a single function $|\Psi_W\rangle$, then density matrix would simply be:
\begin{align}
\rho = |\Psi_W\rangle\langle\Psi_W|
\end{align}
For convenience we introduce the following notation for the basis of Hilbert space:
\begin{align}
\bigg\{|i_1\rangle\otimes\cdots\otimes|i_P\rangle\bigg\} &\longleftrightarrow  \bigg\{|k\rangle\bigg\} \nonumber \\
|\Psi_W\rangle &= \sum_{f} W_{f}|f\rangle.
\end{align}
then the density matrix elements can be written as :
\begin{align}
\rho_{f,f'} = W_kW_{f'} = \rho_{f',f},
\end{align}
which is symmetric due to the real nature of vector space. The above expression works for a pure state, while density matrix itself does not require this, and we can define a general density matrix for arbitrary mixed state as well, then the symmetric nature does not always hold.

For the purpose of later discussion, we introduce reduced density matrix as well. Take a bipartite system consisting of subsystems $A$ and $B$. \textit{If in the Hilbert space this bi-partition corresponds to a well-defined tensor-product structure, i.e. $\mathcal{H} = \mathcal{H}_A \otimes\mathcal{H}_B$}, there would follows a proper basis decomposition, and a state can then be expressed as:
\begin{align}
\bigg\{|i_1\rangle\otimes\cdots\otimes|i_P\rangle\bigg\} &\longleftrightarrow  \bigg\{|g\rangle_A|f\rangle_B\bigg\}, \nonumber \\
|\Psi_W\rangle &= \sum_{g,f} W_{gf}|g\rangle|f\rangle,
\end{align}
where $|g\rangle$ and $|f\rangle$ are proper basis vectors for subsystems $A$ and $B$ respectively. And the associated density matrix is:
\begin{align}
\rho_{gf, g'f'} = W_{gf}W_{g'f'}.
\end{align}
To obtain a reduced density matrix that only depend on the degree of freedom on subsystem $A$, we just need to operate a partial trace on it:
\begin{align}\label{reduced_matrix}
\rho_A &= \sum_{f_B}\langle f |\rho|f\rangle, \nonumber \\
[\rho_A]_{g,g'} &=  \sum_{f} W_{gf}W_{g'f} = [\rho_A]_{g',g},
\end{align}
where we see, that the reduced density matrix from a pure state of the whole system, although in general representing a mixed state of $A$, is still symmetric.

With the precise definition of reduced density matrix, we can now define the von Neumann entanglement entropy~\cite{Eisert_2010}:
\begin{align}\label{vN_EE}
S_{A} = -Tr[\rho_A\log{\rho_A}].
\end{align}
This quantity captures the entanglement between two subsystems $A$ and $B$, and can be reasonably proved to obey the following property:
\begin{align}
S_A = S_B = S_{AB}.
\end{align}
Besides, from the definition \eqref{vN_EE}, we can also obtain two more important results:
\begin{enumerate}
\item Define $M_{A(B)}$ as the dimension of sub-Hilbert space $\mathcal{H}_{A(B)}$, $S_A$ is bounded in the following way:
\begin{align}\label{bound}
S_A \le \log{\max{\{M_A, M_B\}}};
\end{align}
\item $S_A$ is invariant under an arbitrary unitary transformation, i.e.:
\begin{align}\label{basis_free}
-Tr[\rho_A\log{\rho_A}] = -Tr[U\rho_AU^{\dag}\log{U\rho_AU^{\dag}}],
\end{align}
which follows directly from the property of trace operation. This tells us the EE is basis-free.
\end{enumerate}
The bound provided in the first result already suggests a potential connection between the EE and the dimension, which would be discussed further.

One should note that the definition and calculation of EE relies on a partial trace on subsystem $B$. As we mentioned in reduced density matrix introduction, \textit{this would require a proper bi-partition of the whole system, which corresponds to a tensor-product in Hilbert space $\mathcal{H} = \mathcal{H}_A \otimes\mathcal{H}_B$}.

\subsection{Scaling Behaviors of EE}\label{2b}

It is actually very intuitive to connect the entanglement entropy to dimension analysis: since the concept of entropy itself, either in statistical physics or information theory, is related to the number of possible states a system might occupy, which, in the Hilbert space perspective, exactly corresponds to the effective dimension of a problem. Given any function $W$ in the complete Hilbert space and a proper bi-partition, the EE of a subsystem would scale with the size of the subsystem in certain consistent way, i.e. $S_A \sim \mathcal{D}(l_A)$, which is called \textit{scaling behavior of EE}. In this paper, we would use this scaling behavior as the criteria for problem classification.
Now we would list several important scaling behaviors, which corresponds to problems in distinct classes..

For the simplicity of the discussion, we would take a spin-system as an instance to analyze EE scaling. The dimension of sub-Hilbert space $\mathcal{H}_A$ for a spin-$S$ system with in-total $N$ sites is $M_A = (2S+1)^{N_A}$, and for a hyper-cubic shape bi-partition with characteristic length $l_A$, we simply have $M_A=(2S+1)^{l_A^d}$.
Because of the bound \eqref{bound} above, we can obtain the maximal saturation of EE is:
\begin{align}
S_A \le  S_A^{max} \equiv  l_A^d\log{(2S+1)},
\end{align}
which tells us the upper bound saturation $S_A^{max}$ satisfies a so-called \textit{volume law}: 
\begin{align}\label{Smax}
S_A^{max}\sim V_A=l_A^d.
\end{align}
Therefore, we know the EE at most scales linearly with subsystem size.

In fact, given a proper bi-partition of the Hilbert space, most functions would obey a volume law. In other words, if our target function also follows a volume law, then the EE scaling criteria would not effectively reduce the dimension of parameter search (although one may take another completely different criteria in which the concerned problem falls into a minor class). 

On the other hand, however, there are indeed some minority classes in the Hilbert space consisting of much smaller sub-spaces, whose EE scaling is lower than a volume law. In the extreme case, where we can write the whole function into a product state as:
\begin{align}
|\Psi_W\rangle = |\psi^A_W\rangle\otimes|\psi^B_W\rangle,
\end{align}
the reduced density matrix $\rho_A$ (or $\rho_B$) is equivalent to the density matrix of a pure state, whose entanglement entropy is always 0 which can be simply proved as below:
choose a basis of $\mathcal{H}_A$ where $|\psi^A_W\rangle$ is one of its basis vector, then the EE is:
\begin{align}
S_{A} = -Tr[\rho_A\log{\rho_A}] = -\log{1} = 0.
\end{align}
According to \eqref{basis_free}, we know this result is correct in a general basis. In this case, the two subsystems $A$ and $B$ are called \textit{disentangled}, where EE reaches its minimum value. This disentangled class of problem, however, is trivial in the content of optimization, since it corresponds to a separable optimization process, where, in the tensor train decomposition, even virtual indexes $\{\alpha\}$ are not needed at all.

Two nontrivial minor classes are: \textit{area law systems} and \textit{$log$-correction systems}. The so-called area law indicates the scaling of EE for a system in $d$-dimension is:
\begin{align}
S_A \sim l_A^{d-1},
\end{align}
while the EE of a $log$-correction system behaves as:
\begin{align}
S_A \sim l_A^{d-1}\log{l_A}.
\end{align}
The two types of systems defined above have been found in some popular physical systems. It can be proven rigorously, that the EE of a 1d gapped many-body ground state and higher-dimensional conformal field theories (CFT) in general obey the area-law. While the later one, $log$-correction systems, are found for 1d CFTs and free fermion systems in higher dimensions.

Specifically, our RNN model, either in \eqref{PL-prob} or \eqref{spin-wf}, corresponds to a $d=1$ lattice. Now let us analyze $d=1$ in details. In general there are at least four important classes according to their EE scaling behaviors:
\begin{itemize}
\item disentangled system:	 $S_A = 0$;
\item area-law system: $S_A = const >0$;
\item $log$-correction system: $S_A\sim \log{l_A}$;
\item volume-law system: $S_A\sim l_A$.
\end{itemize}

From \eqref{Smax}, which is a general result for any problem, we directly know the upper bound of of our model \eqref{PL-prob} also follows a volume law:
\begin{align}\label{scale0}
S_A\le p_A\log{(L+1)},
\end{align}
where $p_A$ is the number of copies contained in the subsystem $A$ in a certain partition, satisfying $0<p_A<P$. However, the system's true EE scaling behavior is still unknown. The explicit behavior, differs for different dynamics, hence it is not clear here why we choose this criteria for RNN models in HND problems.

\subsection{Tensor Network Representation}\label{2c}

Though we have not explain the explicit connection between the EE scaling criteria and our concerned HND problem, with the chosen criteria in hand, we can construct proper models to match different classes. More precisely, after classifying all functions into different classes according to EE scaling, we can apply models with the same scaling behavior by construction. Then the optimization using a model with explicit scaling behavior would search across only the effective subspace where the true problem lives.

Tensor network (TN) is a widely used representation of generic wavefunctions defined on lattice. In general, the coefficients of a wavefunction in a high dimensional Hilbert space can be viewed as a high-rank tensor. While a tensor network wavefunction decompose the high-rank tensor into lower-rank ones. These lower-rank tensors are usually assigned to lattice sites, which relates the reduced density matrix (hence EE) of a spatial partition to virtual bonds of the local tensors only on the partition boundary. If we allow the virtual bond dimension $D=\infty$, then any higher-rank tensor in principle can be decomposed without errors. However, in practice, one may want to truncate the $D$ to certain finite values, which is followed by some error. Given a fixed error tolerance, not all high-rank tensors can be efficiently decomposed and approximated. Roughly speaking, given a bi-partition of the system, $D$ of virtual bonds on the partition boundary can be related to the entanglement entropy in the following way:
\begin{align}\label{D_S}
D\sim e^{S}.
\end{align}

There are different types of TN. Concerning the purpose of our work, we would introduce two related types of TN used for general 1d systems: namely, Matrix Product State (MPS) and Multiscale Entanglement Renormalization Ansatz (MERA).

\begin{enumerate}
\item MPS:

MPS is the most popular TN used for simulating 1d systems, which has been successfully implemented in DMRG and TEBD methods. Generally, for the coefficient of a 1d wavefunction, one can decompose it into a tensor contraction, with each element given in the following way:
\begin{align}\label{MPS}
W^{i_1i_2\cdots i_P} = \sum_{\{\alpha\}}A^{i_1}_{\alpha_0\alpha_1}A^{i_2}_{\alpha_1\alpha_2}\cdots A^{i_P}_{\alpha_{P-1}\alpha_0},
\end{align}
with each $\alpha$ can take $D$ different possible values, which is also called the virtual bond dimension. 

As revealed by \eqref{D_S}, the efficiency of a TN and system's EE behavior are closely related. There are some systems in condensed matter physics, where the error vanishes exponentially fast when $D$ increases, which makes MPS a good approximation. More precisely, one could describe the efficiency of MPS in two different ways: by examining either the virtual bond dimension $D(L)$ as a function of system size given a fixed error tolerance, or the speed of error vanishing when $D$ increases given the fixed system size. 

In practice, to have a feasible representation of the wavefunction, one would want a constant $D$ that does not increase with the system size if an error bar has been fixed. This request would in general create a special type of TN that intrinsically represents a system whose EE is also a constant: the area-law in $d=1$, which can be easily proved by directly calculating the partial trace of TN wavefunction. If this is the case, then we notice the dimension of the TN model would shrink to:
\begin{align}
M' \propto LD^2 ,
\end{align}
which scales linearly with system size since all local tensors (rank-3) have the same dimension $\propto D^2$. Note that, \textit{reversely, this linear scaling of Hilbert space dimension in 1d would also suggest an area law of EE}.

\item  MERA:

We would briefly mention MERA, and more details can be found in \textit{Evenbly and Vidal 2007}\cite{evenbly} , etc. The general idea of MERA, different from other TNs, is to use a $(d+1)$-dim TN to represent a $d$-dim system, where the extra dimension in physics represents the flow of the Renormalization Group (RG). It has been noticed before that a MERA structure is quite similar to CNN\cite{yahui}.

For 1d cases, MERA describes systems whose EE scales with a $log$-correction, which enters in MERA as a result of spatial coarse-grain in RG. Hence MERA would be a nice candidate for $log$-correction problems, which is more complicated than area-law ones.
\end{enumerate}
We summarize the representing power, by both EE and effective dimension, of different TNs in $d=1$:
\begin{itemize}
\item MPS: disentangled/area-law systems
\begin{align}
S_A = const,\qquad \tilde{M}\sim L;
\end{align}
\item MERA: $log$-correction systems
\begin{align}
S_A \sim \log{l_A}, \qquad \tilde{M}\sim L^{\alpha>1}.
\end{align}
\end{itemize}

Readers may already notice the obvious relation between a MPS \eqref{MPS} and a TT-RNN \eqref{TT-RNN}, which are exactly the same. This is the reason we take a EE perspective, and choose the EE scaling as our criteria. The success of TT-RNN or MPS in the previous work by Yu 2017\cite{yu} suggests that their concerned problems actually live in an effective subspace where all functions inside have an area-law EE scaling, although this has not been realized before.

\section{Hidden Symmetries and Dimension Reduction}\label{3}

We have mentioned the evidence in previous studies that the EE scaling plays an important role. This indicates our chosen criteria is appropriate for analyzing the effective dimension of RNN models in HND problems. With all concepts in mind, in this section, we would directly investigate the model \eqref{PL-prob}, expecting to reduce the effective dimension according to both its intrinsic properties and a problem classification.

More specifically, we would find a hidden symmetry of model \eqref{PL-prob}, which brings in certain gauge redundancy. By fixing gauge and restricting parameters $W$ into a symmetric subspace, the redundancy can be gotten rid of, and effective space has a much lower dimension than before. 

However, in symmetric subspace, a simple bi-partition in original $P$-sites lattice is not proper anymore, since the resulting Hilbert space does not have a tensor-product structure associated with the bi-partition. We raise the question in this section, and would resolve the difficulty in the next one.

\subsection{Gauge Redundancy and Hidden Permutation Symmetry}\label{3a}

The original model \eqref{PL-prob} is a nice representation in practice, since to make $P$ copies of the history information, there is no extra work needed for direct calculation of higher order polynomials. However, the fact that we are only looking at an order-$P$ lag-$L$ problem makes the ordering in the sequence $\{i_1\cdots i_P\}$ meaningless. For example, considering the following two states (functions):
\begin{align}
|h_{t-1}\rangle\otimes|1\rangle\otimes\cdots\otimes|1\rangle, \nonumber\\
|1\rangle\otimes\cdots\otimes|1\rangle\otimes|h_{t-1}\rangle,
\end{align}
where only one element takes the value $h_{t-1}$ while all other taking constant $1$. In the original Hilbert space setting \eqref{Hilbert0}, these two states are orthogonal basis functions, and the coefficients of them would be treated independently. However, since we only care about independent polynomials, there is a redundancy that would produce the same result:
\begin{align}
\lambda_1|h_{t-1}, 1 \cdots 1\rangle + \lambda_P| 1 \cdots 1, h_{t-1}\rangle, \quad s.t. \text{ }\lambda_1+\lambda_P = \Lambda.
\end{align}
And we can put $h_{t-1}$ in arbitrary position, and obtained:
\begin{align}
\sum_{m=1}^{P}\lambda_m|\underbrace{1\cdots 1}_{m-1}, h_{t-1}, \underbrace{1\cdots 1}_{P-m}\rangle, \quad s.t. \text{ } \sum\lambda_m = \Lambda.
\end{align}
Any combination satisfying $\sum\lambda_m= \Lambda$ would produce the same polynomial. 

Most generally, given a sequence:
\begin{align}
|i_1\rangle\otimes\cdots\otimes|i_P\rangle, \quad \forall i\in \big[1, h_{t-1}, \cdots, h_{t-L}\big],
\end{align}
the only thing matters for composing a polynomial is the number of times each $h_{\tau}$ appears. This in general is called a \textit{gauge redundancy} in the context of mathematics and theoretical physics. In the language of coefficients $W^{\{i\}}$, there exists a transformation to relate two equivalent coefficients:
\begin{align}
\tilde{W}^{i'_1i'_2\cdots i'_P} = \hat{O}_{i_1i_2\cdots i_P}^{i'_1i'_2\cdots i'_P} W^{i_1i_2\cdots i_P}, 
\end{align}
with a default Einstein summation on all dummy indexes; or, more concisely:
\begin{align}
\tilde{W} = \hat{O}\circ W,
\end{align}
where $\hat{O}$ represents the transformations. And the transformed wavefunction $\psi_{\tilde{W}}$, though representing a different state in the original Hilbert space, preserves all properties as a polynomial used in real calculations.

When taking the perspective of the original Hilbert space, the existence of gauge redundancy always suggests certain \textit{symmetries} in the concerned space. Mathematically, all these symmetry operations would form a symmetry group, and one could use the group representation theory to construct appropriate coefficients $W$. The symmetry associated with gauge redundancy mentioned above is actually a \textit{permutation symmetry} on all $P$ indexes, and the resulting group is the \textit{permutation group} $S_{P}$. This symmetry, would provide a powerful tool for theoretical analysis below on the truly desired subspace.

\subsection{Dimension Reduction and Projected Subspace}\label{3b}

With the redundancy and the associated $S_P$ symmetry, we could restrict our discussion in a smaller efficient subspace: $\mathcal{H}_0$.

Now let us calculate explicitly the dimension of $\mathcal{H}_0$. This is equivalent to a polynomial problem: the dimension of $\mathcal{H}_0$ equals to the number of independent $L$-multivariate polynomials upto $P$-th order, which can be simply derived by adding an extra constant $1$ as another variable, and the result is:
\begin{align}\label{dim1}
\tilde{M} = C_{L+P}^{P} = \frac{(L+P)!}{L!P!}.
\end{align}
This is much smaller than the original Hilbert space dimension: $M=(L+1)^P$. 

Theoretically, we can find $\mathcal{H}_0$ by fixing a gauge in the original representation, a typical procedure used in gauge theory. For simplicity, we would use the following gauge:
\begin{align}
\tilde{W}^{\bar{\{i\}}} \equiv W^{\{i\}} = W^{\pi\circ\{i\}},
\end{align}
where $\pi\circ\{i_1i_2\cdots i_P\}$ represents an arbitrary permutation of $\{i_1i_2\cdots i_P\}$, which is an element in the group $S_P$. A proper basis vector for the current subspace would then be the following symmetric combination:
\begin{align}
\overline{i_1i_2 \cdots i_P} = \frac{1}{K}\sum^K_{\pi\in S_P}\pi\circ\{i_1i_2\cdots i_P\},
\end{align}
an equal-weight average of all the permutation-accessible original vector, where $K$ is the rank of the permutation group. In this symmetric basis,
\begin{align}
\overline{i_1i_2 \cdots i_P} \equiv \overline{\pi\circ i_1i_2 \cdots i_P}
\end{align}
represents the same state with different labeling, other than different states. This is a typical gauge theory as discussed in theoretical physics. Each element in this $\tilde{M}$-dim Hilbert space is invariant under the group action of $S_P$. Therefore we have now:
\begin{align}
\mathcal{H}_0: \bigg\{|\overline{i_1i_2 \cdots i_P}\rangle\bigg\}, \quad \forall i\in \big[1, h_{t-1}, \cdots, h_{t-L}\big].
\end{align}

The original representation of the problem \eqref{PL-prob}, however, has the advantage for practical computation. To fill the gap between theoretical analysis and the practically beneficial model, we simply connect the two Hilbert space by a projection operator:
\begin{align}
\mathcal{P}\circ \mathcal{H} = \mathcal{H}_0,
\end{align}
then any calculation can be firstly analyzed in the original model, and then be projected onto subspace $\mathcal{H}_0$.

We now discuss the impact of the permutation symmetry on a subsystem $A$ in certain bi-partition. Firstly, viewed individually, the dimension of sub-Hilbert space is:
\begin{align}
\tilde{m}_A = C_{L+p_A}^{L} = \frac{(L+p_A)!}{L!p_A!},
\end{align}
at the same time, for the subsystem $B$ we similarly have:
\begin{align}
\tilde{m}_B = C_{L+p_B}^{L} = \frac{(L+p_B)!}{L!p_B!},
\end{align}
where $p_{A/B}$ satisfying $p_A  + p_B = P$. These dimensions for subsystems is already much smaller than the upper bound which follows a volume law as in \eqref{scale0}. Therefore, at least we would expect a sub-volume law for the EE scaling. In the next chapter, using a different representation of the same problem, we could further talk about EE scaling classification of different HND systems.

On the other hand, this individual dimension counting, however, is still more than necessary, due to the overlook of mutual permutations between the two subsystems, which can be shown from the following inequality:
\begin{align}
C_{L+p_A}^{L}\cdot C_{L+p_B}^{L} \geq C_{L+p_A+p_B}^{L},
\end{align}
or in the group language:
\begin{align}
S_{p_A}\times S_{p_B} \subseteq S_{p_A+p_B}.
\end{align}
This would further reduce the Hilbert space dimension of subsystem $A$ and EE between $A$ and $B$.

\subsection{Ambiguity in Entanglement Entropy Calculation}\label{3c}
The above symmetric subspace $\mathcal{H}_0$ not only simplify the parameter searching, but also is more reasonable as a polynomial approximation. It is a general analysis for any RNN models in HND problems as \eqref{PL-prob}. 
As promised, to progress further, we would use the previous introduced EE scaling criteria to classify possible problems and match them with proper models. However, there is a problem arising in our current case.

The symmetry property helps us to restrict the problem into a sub-Hilbert space which has a much lower dimension than the original one. This basis change on the whole system itself is not a problem, but it brings a big trouble when we look at any subsystem. 
As we mentioned before, an unambiguous definition of EE requires a proper partition of the whole system; and a partition is a proper one only if it produces a tensor-product decomposition of the whole Hilbert space:
\begin{align}
\mathcal{H} = \mathcal{H}_A \otimes \mathcal{H}_B.
\end{align}
While in our present case, if we choose a partition into $p_A$-copies and $p_B$-copies, being local permutation symmetric does not promise an overall-permutation symmetry. Thus if the whole Hilbert space $\mathcal{H}_0$ is overall-permutation symmetric, then we cannot find proper local basis functions of subsystems $A$ and $B$ to accommodate the tensor-product structure $\mathcal{H}_A \otimes \mathcal{H}_B$.

\section{Dual Representation and Entanglement}\label{4}
In the last section, after reducing the dimension of Hilbert space by taking account into the hidden permutation symmetry, we are supposed to further discuss the explicit EE scaling behavior to implement models like TT-RNN. However, due to symmetric nature of the reduced parameter space, the original model itself does not accommodate a proper bi-partition of the whole Hilbert space. 

To resolve this issue and find a rigorous way to calculate (or define) EE in an arbitrary problem, we would introduce a new representation of the original model. In the above section we proved that the searching-problem actually lives in a subspace $\mathcal{H}_0$, whose dimension scales as:
\begin{align}
\frac{(L+P)!}{L!P!}.
\end{align}
The expression is symmetric w.r.t $L$ and $P$. This suggests there should be another way to construct the model, at least theoretically (may not be a feasible one in practice as explained below), to flip the roles of $P$ and $L$. Inspired by this symmetric expression, we introduce a dual representation with a new Hilbert space, which, also as a redundant one, contains a same subspace as the original one where the searching-problem lives. With the assistance of this representation, we are able to apply the proposed criteria for problem classification in the original model representation. 

\subsection{Dual Representation of Projected Subspace}\label{4a}

We introduce another 1d lattice system, which contains $L$ sites, and the local Hilbert space dimension constantly equals $P$. We would call it the \textit{time-lag space}. Formally, all available local states on a site $r$ are:
\begin{align}
[h_{t-r}^0, h_{t-r}^1, h_{t-r}^2, \cdots, h_{t-r}^P]\qquad \forall r\in[0, L].
\end{align}
where for simplicity we abuse the notation and redefine:
\begin{align}
h_{t} \equiv 1.
\end{align}
The dual version of the problem can be written as:
\begin{align}\label{dual-prob}
\phi_V &= tTr\bigg(V^{j_0\cdots j_L}q_{j_0}\otimes\cdots\otimes q_{j_L}\bigg), \nonumber \\
q_{j_r}&\in [h_{t-r}^0, h_{t-r}^1, h_{t-r}^2, \cdots, h_{t-r}^P],
\end{align}
which, in the language of wavefunction, can be formulated as:
\begin{align}\label{dual-wf}
|\Phi\rangle_V = \sum_{\{j_0, \cdots, j_L\}} V^{j_0\cdots j_L} |j_0\cdots j_L\rangle, \qquad s.t. \bigg(\sum_{\tau}j_{\tau} = P\bigg).
\end{align}

Firstly, the Hilbert space dimension of this dual model is 
\begin{align}
(P+1)^{L+1},
\end{align}
which, similar as before, is redundant. The upper bound of an arbitrary partition would behave as:
\begin{align}\label{dual_uppper}
S_{A} \leq (l_A+1)\log{(P+1)},
\end{align}
which follows a volume law in the sense of a partition in the time-lag space, where $l_A$ is the length of subsystem $A$.

Secondly, in the case $P>L$ (or limit $P\gg L$), the dual model in practice behaves better than the original one if we just consider the searching process of parameters $V$.

Thirdly, this space, as a redundant one, is a legal choice because it contains the same sub-Hilbert space where the real problem lives. Recall that we actually only need all $L$-multivariate polynomials upto order $P$, while the basis functions as below:
\begin{align}
\bigg\{|j_0\rangle\otimes\cdots\otimes |j_L\rangle\bigg\},
\end{align}
besides polynomials upto order $P$, can compose even higher order ones such as:
\begin{align}
1^Ph_{t-1}^Ph_{t-2}^P\cdots h_{t-L}^P,
\end{align}
which is of order $LP$. Hence it is a legal but also redundant space.

The expanding dimensions of the symmetric subspace $\mathcal{H}_0$ defined previously, in the current case, have a clearer geometric meaning: \textit{they form a "hyper-plane" $\mathcal{D}_{\mathcal{H}_0}$ in the whole "dimensionality space" $\mathcal{D}_{\mathcal{H}}$}. For example, let us take $L=1$ and $P=6$ for an example, i.e. we have two "sites": $h_{t}\equiv 1$, and $h_{t-1}$. Heuristically, $\mathcal{H}_0$ can be shown in Fig.\ref{fig_dim_hilberr}:
\begin{figure}[!ht]
\includegraphics[width = 7cm]{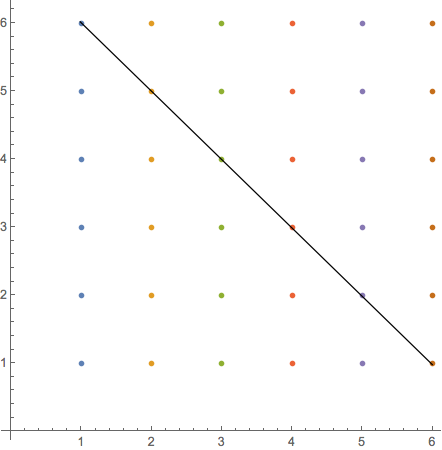}
\caption{Effective dimensions form a "hyper-plane". Horizontal axis: the power of $h_t\equiv1$; Vertical axis: the power of $h_{t-1}$.}
\label{fig_dim_hilberr}
\end{figure}

In the figure above, each single point represent an expanding dimension (a basis function) of the whole Hilbert space. While the whole space consists of 36 points forming a square shape on the $h_t$-$h_{t-1}$ plane, due to the restriction of the total polynomial order, the allowed subspace would be a line segment. When $L$ is larger, this line segment expands into a hyper plane in the higher dimensional ($d=L+1$) space.

\subsection{Entanglement Entropy and Dimension Analysis}\label{4b}

We mentioned in the last section that original representation after symmetrization is not a proper choice for entanglement entropy calculation, and hence does not provide a playground for dimension analysis based on EE behavior classification. And the dual model can help to resolve this puzzle. 

The nice feature we have in the dual representation is that the symmetric Hilbert space $\mathcal{H}_0$ now has an explicit tensor product structure, because no further symmetrization is required anymore:
\begin{align}
\mathcal{H}_0 = \mathcal{H}_{A,0} \otimes \mathcal{H}_{B,0},
\end{align}
where the subsystems $A$ and $B$ represent a partition in the time-lag space: a 1d chain with $L$ "sites". And we could directly choose the local basis for subsystems as:
\begin{align}
|g\rangle= |j_0\cdots j_{l_A}\rangle, \qquad |f\rangle= |j_{l_A + 1}\cdots j_{l_A+l_B\equiv L}\rangle, 
\end{align}
which then gives an EE expression:
\begin{align}
S_A &= -Tr\big[\rho_A\log{\rho_A}\big] \nonumber \\
&= -\sum_{g_A} \langle g|\rho_A\log{\rho_A}|g\rangle, \nonumber \\
\rho_A &= \sum_{f_B}\langle f |\rho|f\rangle, \qquad [\rho_A]_{g,g'} =  \sum_{f} V_{gf}V_{g'f}.
\end{align}
Therefore, in this representation, we can discuss the scaling behavior of EE and the resulting effective dimension. 

As in general $d=1$ systems, there would be majorly four different classes: disentangled, area-law, $log$-correction, volume law. For the last one, a model construction w.r.t the current criteria is not helpful. For the other three, one can use tensor-network models to approximate the system effectively. Now since we obtain the EE scaling and classification in this dual model, a model construction and followed computation should in principle also be implemented in this dual representation \eqref{dual-prob} -- this is troublesome since it requires a direct calculation of higher order terms, bringing extra computation cost than the original one \eqref{PL-prob}. 
Besides, our attempt to explain the existing success of TT-RNN in previous work\citep{yu} cannot be accomplished in this dual representation either. How could one relate the result in the dual representation to the original basis used in previous works?

The answer is about the effective dimension, which, given a question and a classification criteria, should be understood as the number of "degree of freedom", or say, the number of \textit{independent parameters}. This is an intrinsic property coded in the problem itself other than any basis choices. We have already proved that the two representing spaces contain the same target subspace. If we then found the effective dimension of a certain problem in the dual basis, it should also be applicable in the original basis.

\begin{figure}[!ht]
\centering
\includegraphics[width=8cm]{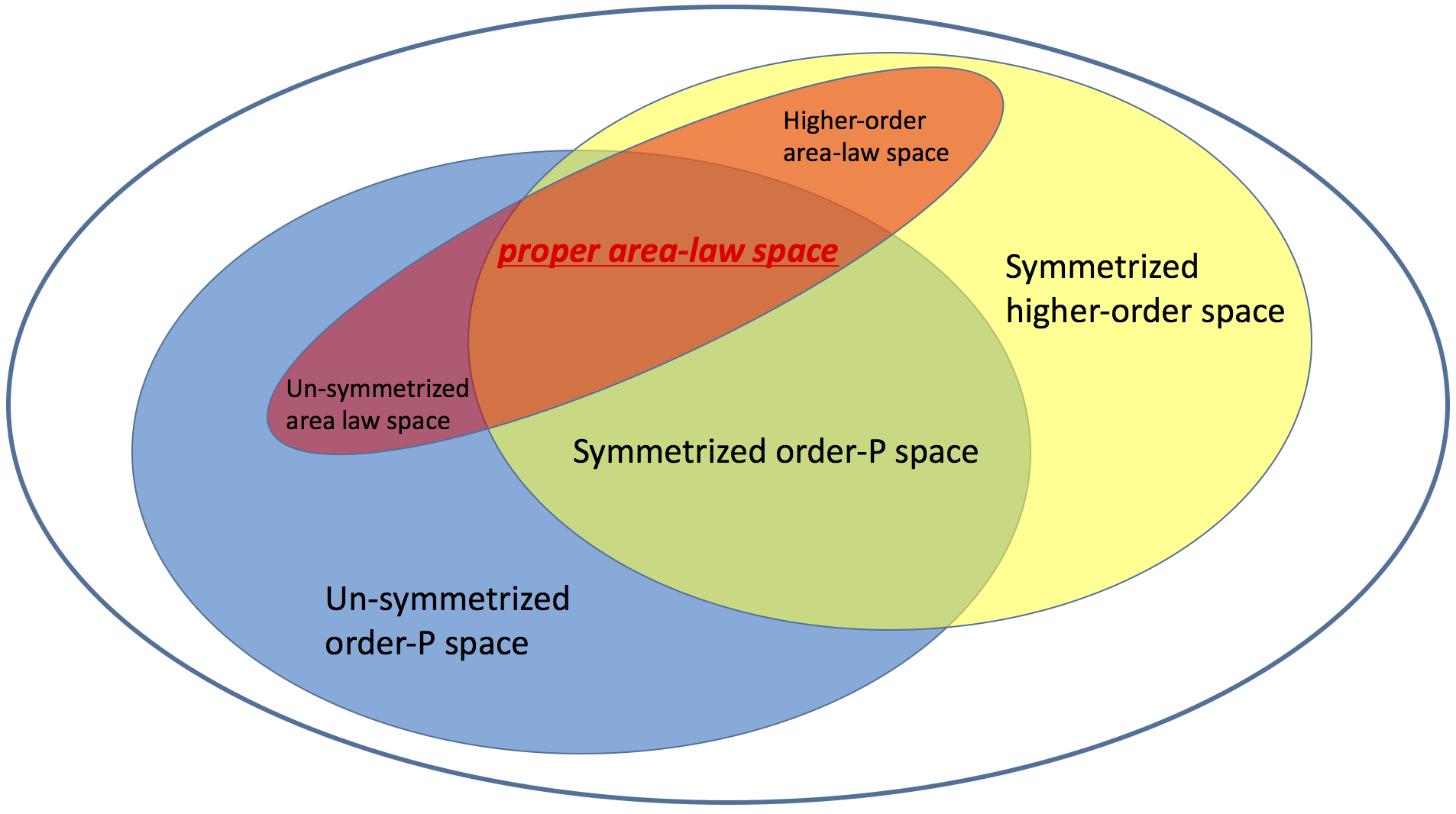}
\caption{Relation of different sub-Hilbert-space}
\label{fig_Hilbert_relation}
\end{figure}

Let us analyze area-law systems as an example: given an order-$P$ lag-$L$ problem whose EE, in the dual basis, obeys an area law, then from \eqref{D_S} we know the effective dimension of the system in the dual representation is:
\begin{align}\label{eff_dim}
\tilde{M}  = D^2 LP.
\end{align}
Since the effective dimension of a problem in the given EE-scaling criteria is an intrinsic property, this result also applies to the original model representation \eqref{PL-prob}. 
To be more heuristic, the relation among different parts of the whole Hilbert space are shown in Fig.\ref{fig_Hilbert_relation}.
Note the three area law regions have different meanings of "area-law", while we just for simplicity plot them continuously.

Now going back to the original basis with $P$-replicas of the history, with or without the symmetrization, \eqref{eff_dim} ensures that there exists a tensor decomposition structure as:
\begin{align}\label{mps_ttrnn}
W^{i_1i_2\cdots i_P} = \sum_{\{\alpha\}}A^{i_1}_{\alpha_0\alpha_1}A^{i_2}_{\alpha_1\alpha_2}\cdots A^{i_P}_{\alpha_{P-1}\alpha_P},
\end{align}
where each virtual index $\alpha$ lives in a $D$-dim local space. This is exactly a MPS representation, or a TT-RNN model, whose total dimension scales as \eqref{eff_dim}.

In another words, even after incorporating the symmetric nature of a parameter space, we can still use EE scaling criteria to classify problems. And therefore, the TT-RNN structure would still be a valid choice.

\section{A More Physical Perspective: Identical Particle System}

By taking both original and dual representations, we are able to clarify the EE definition and hence the proposed classification criteria, and justify the legitimacy of implementing TT-RNN in HND systems. While mathematically rigorous, the above discussion does not provide a physically intuitive picture, despite the EE itself has a clear physical meaning.

In this section, we would map the system into an identical boson system. We would firstly formulate this mapping, and discuss the relation to previous mutually dual representations. Then from the resulting perspective of a generic identical particle system, we would rephrase the EE scaling and area law by an analogy to quantum Hall systems.

\subsection{Mapping to an Identical Boson System}\label{4c}

The existence of the permutation symmetry reminds us the Bose-Einstein statistics. Indeed, there is a deep connection between the current models with an identical boson system. We could consider a $P$-particles bosonic system with the following \textit{$(L+1)$ available single particle states}:
\begin{align}
h_t, h_{t-1}, \cdots , h_{t-L}.
\end{align}
Again, we define here $h_t\equiv 1$. Different single particle states can be viewed as orthogonal eigenstates of certain "Hamiltonian" operator $H$.

Now the whole system is a bosonic many-body system, and we have two ways to write down the wavefunction: in either the first or the second quantization languages.

In the first quantization language, a wavefunction would be one with a symmetrization of all $P$ particles:
\begin{align}\label{first}
|\Psi\rangle_W &= \sum_{\overline{i_1\cdots i_P}} W^{\overline{i_1\cdots i_P}}\bigg[\frac{1}{K}\sum^K_{\pi\in S_P}|\pi\circ i_1\cdots i_P\rangle\bigg] \nonumber \\
&= \sum_{\overline{i_1\cdots i_P}} W^{\overline{i_1\cdots i_P}}|\overline{i_1\cdots i_P}\rangle,
\end{align}
where, as our convention before, $\overline{i_1\cdots i_P}$ represents symmetrized sequences. We immediately notice this is similar as \eqref{PL-prob} and \eqref{spin-wf}, just with an extra symmetrization.

Now we go into the second quantization language. We could define the associated creation/annihilation operators for each single particle state as:
\begin{align}\label{single}
b^{\dag}_{\tau}|0\rangle_{\tau} = |h_{t-\tau}\rangle, \qquad \forall \tau\in[0,L],
\end{align}
where $|0\rangle_{\tau}$ is the vacuum of certain single particle, which in the polynomial content always equals $1$. They follow the usual bosonic commutation relations:
\begin{align}
\big[b_{\tau}, b_{\tau'}\big]= \big[b^{\dag}_{\tau}, b^{\dag}_{\tau'}\big] =0, \qquad \big[b_{\tau}, b^{\dag}_{\tau'}\big] = \delta_{\tau\tau'}.
\end{align}
With this in hand, the whole many-body wavefunction is then simply:
\begin{align}\label{second}
|\Phi\rangle_V = \sum_{\{j_0, \cdots, j_L\}} V^{j_0\cdots j_L} |j_0\cdots j_L\rangle, \qquad s.t. \bigg(\sum_{\tau}j_{\tau} = P\bigg),
\end{align}
where $|j_0\cdots j_L\rangle$ is an eigenstate of occupation number operators, constructed as below:
\begin{align}
|j_0\cdots j_L\rangle = \prod_{\tau=0}^L\frac{(b_{\tau}^{\dag})^{j_{\tau}}}{\sqrt{j_{\tau}!}}|0\rangle.
\end{align}
And $|0\rangle$ represents the vacuum state of the many-body system: 
\begin{align}
|0\rangle = \prod_{\otimes\tau}|0\rangle_{\tau}.
\end{align}
This form now is completely the same as \eqref{dual-prob} and \eqref{dual-wf}.

Therefore, taking an identical boson perspective, the relation between two representation is quite clear. The two different bi-partitions for EE calculation in two representations then correspond to the so-called \textit{particles bi-partition} and \textit{modes bi-partition} \citep{dalton, benatti} in the entanglement study of identical particle systems, which is a well-studied problem in quantum information and fundamentals of quantum mechanics.

\subsection{EE scaling and Quantum Hall Systems}

For people from quantum information, it is rare to study EE scaling behavior in identical particle systems. This, however, has been discussed in details in Quantum Hall Effect (QHE) ~\cite{orus}. 

An important property (especially for EE calculation) of QHE systems is the itinerant nature of electrons, which differs from spin system with each spin sitting on the one site. For EE calculation, CFTs actually pertain the blocking of space rather than particles, which turns out to be not a natural description in numeric calculations of QHE. In recent studies, two popular ways of partition are used: orbital partition and particle partition \cite{haque, zozulya}:
\begin{enumerate}
\item \textbf{Orbital partition }corresponds to a partition of Landau levels into two parts. Landau levels actually describe single particle states, and have large degeneracy. For QHE systems in a symmetric gauge, orbital partition is closely related to spatial
partition since Landau orbits form disk-like region in real space that has an exponential decay of spatial expansion.

\item \textbf{Particle partition} corresponds to a direct partition of particles in the whole system, which in QHE does not
corresponds to a connected spatial region at all. 
\end{enumerate}
The orbital partition is actually one kind of modes partition in the second quantization picture, while the particle partition is the same as the previous discussion in the first quantization picture. 

Although both orbital partition and particle partition in QHE can be argued to be related to spatial partition, the itinerant nature of electrons and hence the extended nature of wavefunctions actually still make a difference . However, one can still talk about the EE scaling behavior with a change of measure, from spatial size to particle numbers, which is used in particle partition method \cite{haque}. This is a proper perspective to explain EE scaling in both the original model \eqref{PL-prob} and first quantization wavefunctions \eqref{first}.

\section{General Tensor Network Structures for Sequential Modeling}\label{6}
In this section, we would provide several classes of TN structures for general sequential modeling tasks. Reminded that we have discussed two representations and two TN structures, we propose the following models.

\subsection{MPS in two representations}

One way to incorporate higher order (upto $P$) terms, is to make $P$ identical copies of the delayed state variable, and use their product. A general model looks like:
\begin{align}
x_{t} &= F\bigg[\sum_{\{l_1, \cdots, l_P\}}W^{l_1 \cdots l_P}\cdot (x_{t-l_1}x_{t-l_2}\cdots x_{t-l_P})\bigg], \qquad l_i = \{0, \cdots L\},
\end{align}
where on the RHS we for convenience abuse the symbol by redefining: $x_{t-0} \equiv 1$.
The dual version of the model would be:
\begin{align}
x_{t} &= F\bigg[\sum_{\{p_1, \cdots, p_L\}}V^{p_1 \cdots p_L}\cdot (x_{t-1}^{p_1}x_{t-2}^{p_2}\cdots x_{t-L}^{p_L})\bigg], \qquad p_j = \{0, \cdots P\}.
\end{align}
Implement a MPS representation (approximation):
\begin{align}
W^{l_1 \cdots l_P} &\simeq A_{r_0r_1}^{l_1}A_{r_1r_2}^{l_2}\cdots A_{r_{P-1}r_P}^{l_P} \\
V^{p_1 \cdots p_L} &\simeq A_{i_0i_1}^{p_1}A_{i_1i_2}^{p_2}\cdots A_{i_{L-1}i_L}^{p_L}
\end{align}
This is the TT-RNN model. It works only when hidden area-law exists in the system.

\subsection{MERA in dual rep}

For more complicated cases where MPS does not work, a MERA would be helpful. 

In the dual representation, take an example of $L=4$, the MERA structure (periodic) for a wave-function would be:
\begin{align}
V^{p_1p_2p_3p_4} &\simeq \sum_{\{q_1, q_2\}}\tilde{\tilde{V}}^{q_1q_2}\tilde{V}_{q_1}^{r_1r_2}\tilde{V}_{q_2}^{r_3r_4}\hat{V}_{r_1r_2}^{p_4p_1}\hat{V}_{r_3r_4}^{p_2p_3}.
\end{align}
Not rigorous but innocuous to our purpose, we can for now ignore the dimension-keeping disentangler $\hat{V}$, and just write down:
\begin{align}
V^{p_1p_2p_3p_4} &\simeq \sum_{\{q_1, q_2\}}\tilde{\tilde{V}}^{q_1q_2}\tilde{V}_{q_1}^{p_1p_2}\tilde{V}_{q_2}^{p_3p_4}.
\end{align}
Then the whole wave-function becomes:
\begin{align}
\psi &= \sum_{\{p_1p_2p_3p_4\}}\sum_{\{q_1, q_2\}}\tilde{\tilde{V}}^{q_1q_2}\tilde{V}_{q_1}^{p_1p_2}\tilde{V}_{q_2}^{p_3p_4} x_{t-1}^{p_1}x_{t-2}^{p_2}x_{t-3}^{p_3}x_{t-4}^{p_4}.
\end{align}
For convenience, we rewrite the wave-function as:
\begin{align}
y_{1, q_1} &= \sum_{\{p_1p_2\}}\tilde{V}_{q_1}^{p_1p_2}x_{t-1}^{p_1}x_{t-2}^{p_2} \nonumber\\
y_{2, q_2} &= \sum_{\{p_3p_4\}}\tilde{V}_{q_2}^{p_3p_4}x_{t-3}^{p_3}x_{t-4}^{p_4} \nonumber\\
\psi &= \sum_{\{q_1, q_2\}}\tilde{\tilde{V}}^{q_1q_2}y_{1, q_1}y_{2, q_2}
\end{align}
To accommodate a sequence modeling problem, we operate a nonlinear function on all meta-variables, and obtain:
\begin{align}
y_{1, q_1} &= F\bigg[\sum_{\{p_1p_2\}}\tilde{V}_{q_1}^{p_1p_2}x_{t-1}^{p_1}x_{t-2}^{p_2}\bigg] \nonumber\\
y_{2, q_2} &= F\bigg[\sum_{\{p_3p_4\}}\tilde{V}_{q_2}^{p_3p_4}x_{t-3}^{p_3}x_{t-4}^{p_4}\bigg]  \nonumber\\
x_t &= F\bigg[\sum_{\{q_1, q_2\}}\tilde{\tilde{V}}^{q_1q_2}y_{1, q_1}y_{2, q_2}\bigg] 
\end{align}
Now let us find the relation to a TCN model by putting some constraints. For example, the following choice would work:
\begin{align}
\tilde{V}_{q_1}^{p_1p_2} &= \delta_{q_1}^1\cdot\bigg(\delta^{p_1}_1\delta^{p_2}_0\bar{V}_1^1 + \delta^{p_1}_0\delta^{p_2}_1\bar{V}_1^2\bigg) + \delta_{q_1}^0\delta^{p_1}_0\delta^{p_2}_0\Lambda, \nonumber \\
\tilde{V}_{q_2}^{p_3p_4} &= \delta_{q_2}^1\cdot\bigg(\delta^{p_3}_1\delta^{p_4}_0\bar{V}_2^1 + \delta^{p_3}_0\delta^{p_4}_1\bar{V}_2^2\bigg) + \delta_{q_2}^0\delta^{p_1}_0\delta^{p_2}_0\Lambda, \nonumber \\
\tilde{\tilde{V}}^{p_1p_2} &= \delta^{q_1}_1\delta^{q_2}_0\bar{\bar{V}}^1 + \delta^{q_1}_0\delta^{q_2}_1\bar{\bar{V}}^2,
\end{align}
where $\Lambda$ is some constant large enough, such that given an error bar $\epsilon$, we have $1 - F[\Lambda] < \epsilon$, and therefore $y_{1,0}=y_{2,0}\simeq 1$.
Applying this set-up, the whole model would be equivalent to:
\begin{align}
y_1 &= F\bigg[\bar{V}_1^1x_{t-1} + \bar{V}_1^2x_{t-2}\bigg], \nonumber\\
y_2 &= F\bigg[\bar{V}_2^1x_{t-3} + \bar{V}_2^2x_{t-4}\bigg], \nonumber\\
x_t &=  F\bigg[\bar{\bar{V}}^1y_1 + \bar{\bar{V}}^2y_2\bigg].
\end{align}
This is exactly a 2-layer TCN structure. 
Importantly, in TCN paper, they used the same filter $\bar{V}$ across a layer, i.e. in the above expression requiring $\bar{V}_1 = \bar{V}_2$. This, though reduce the computational cost, is not a reasonable setting, since the history impact from different moments should be treated equally.

\subsection{MERA in original representation}

In the original representation, take an example of $P=4$, the MERA structure (periodic) for a wave-function would be:
\begin{align}
W^{l_1l_2l_3l_4} &\simeq \sum_{\{m_1, m_2\}}\tilde{\tilde{W}}^{m_1m_2}\tilde{W}_{m_1}^{l_1l_2}\tilde{W}_{m_2}^{l_3l_4}.
\end{align}
Then the whole wave-function becomes:
\begin{align}
\psi &= \sum_{\{l_1l_2l_3l_4\}} \sum_{\{m_1, m_2\}}\tilde{\tilde{W}}^{m_1m_2}\tilde{W}_{m_1}^{l_1l_2}\tilde{W}_{m_2}^{l_3l_4} \cdot(x_{t-l_1}x_{t-l_2}x_{t-l_3}x_{t-l_4}).
\end{align}
And we can obtain the sequence model as the following:
\begin{align}
y_{1, m_1} &= F\bigg[\sum_{\{l_1l_2\}}\tilde{W}_{m_1}^{l_1l_2}x_{t-l_1}x_{t-l_2}\bigg] \nonumber\\
y_{2, m_2} &= F\bigg[\sum_{\{l_3l_4\}}\tilde{W}_{m_2}^{l_3l_4}x_{t-l_3}x_{t-l_4}\bigg]  \nonumber\\
x_t &= F\bigg[\sum_{\{m_1, m_2\}}\tilde{\tilde{W}}^{m_1m_2}y_{1, m_1}y_{2, m_2}\bigg] 
\end{align}
Similarly, by choosing special form of the tensors, it could also produce the TCN model:
\begin{align}
\tilde{W}_{m_1}^{l_1l_2} &= \delta_{m_1}^1\cdot\bigg(\delta^{l_1}_1\delta^{l_2}_0\bar{V}_1^1 + \delta^{l_1}_0\delta^{l_2}_2\bar{V}_1^2\bigg) + \delta_{m_1}^0\delta^{l_1}_0\delta^{l_2}_0\Lambda, \nonumber \\
\tilde{W}_{m_2}^{l_3l_4} &= \delta_{m_2}^1\cdot\bigg(\delta^{l_3}_3\delta^{l_4}_0\bar{V}_2^1 + \delta^{l_3}_0\delta^{l_4}_4\bar{V}_2^2\bigg) + \delta_{m_2}^0\delta^{l_1}_0\delta^{l_2}_0\Lambda, \nonumber \\
\tilde{\tilde{W}}^{m_1m_2} &= \delta^{m_1}_1\delta^{m_2}_0\bar{\bar{V}}^1 + \delta^{m_1}_0\delta^{m_2}_1\bar{\bar{V}}^2,
\end{align}
Again, requiring $\bar{V}_1 = \bar{V}_2$ seems not to be a reasonable simplification.

However, without the above special choice, due to the redundancy from permutation symmetry, in this original representation we indeed can use the same filter across a single layer, which, in this representation, requires:
\begin{align}
\tilde{W}_{m}^{l_al_b} &= \tilde{W}_{m'}^{l_bl_a}, \qquad \forall a,b,\forall m,m' \nonumber \\
\tilde{\tilde{W}}^{mm'} &= \tilde{\tilde{W}}^{m'm}, \qquad \forall m,m'
\end{align}
The symmetry of the lower index $m$ in $\tilde{W}$s represents the same filter in the first layer. The symmetry of all upper indexes comes from the permutation symmetry.

\section{Conclusion and Discussion}\label{7}
Inspired by the previous proposal of a higher order tensor-train RNN(TT-RNN) model, we established a criteria for the problem classification within the higher order RNN formulation. This criteria classifies problems according to their entanglement entropy(EE) scaling behaviors. The success of TT-RNN model in dimension reduction is therefore rooted in certain area-law of EE scaling of the concerned problems themselves. 

To further reduce the dimension of the parameter space and hence computational cost, we discovered a hidden permutation symmetry which results in a huge gauge redundancy. While a smaller symmetric sub-space is legal for a practical computation, the EE calculation in any symmetric space becomes vague, which ruins the criteria for the problem classification and makes the TT-RNN structure inapplicable.

The discussion of EE in symmetric spaces is legitimated by introducing a dual representation, which recovers a proper definition of EE. The problem classification in two representations are connected through an argument of effective dimension. Therefore, the TT-RNN structure becomes valid even in a symmetric parameter space. And we derive the symmetric form of a TT-RNN model in the end, which in practice should reduce the computational cost dramatically.

Besides, we map the high-order RNN problem to an identical boson system, and we also find a relation of the partition of our system and the partition used in QHE numeric calculations, which would bring more intuition for EE analysis.  

The proposed symmetric form of a TT-RNN model could be directly implemented in numeric practice. As a time-series modeling, an more interesting question to be asked is the relation between the entanglement entropy and the time-correlation. Since the difficulty of describing a dynamical systems can be characterized by both the EE scaling and the chaotic nature, it would be extremely interesting if these two criteria can be related. On the other hand, within our proposed criteria, another important class is the so-called $log$-correction systems, where a TT-RNN is not a proper choice anymore, and hence more complicated models are desired for these systems. In another on-going work, still starting from a high-order RNN structure, we proposed a new model using Multi-scale Entanglement Renormalization Ansatz. The new proposed model would provide an unified perspective for the two most popular sequence modeling structures: the RNN structure and the CNN structure.

\bibliography{draft2Notes}

\end{document}